# Efficient and Sustainable Treatment of Tannery Wastewater by a Sequential Electrocoagulation-UV Photolytic Process

Jallouli Sameh[1]*, Ahmed Wali[1], Antonio Buonerba[2,3], Tiziano Zarra[2,3], Vincenzo Belgiorno[2,3], Vincenzo Naddeo[2,3]* and Mohamed Ksibi[1]

[1] Université de Sfax, Laboratoire de Génie de l'Environnement et Ecotechnologie, GEET-ENIS, Route de Soukra km 4 Po. Box 1173, Sfax 3038, Tunisia.

[2] Sanitary Environmental Engineering Division (SEED), Department of Civil Engineering, University of Salerno, Via Giovanni Paolo II, 84084 - Fisciano (SA), Italy.

[3] Inter-university Consortium for Prediction and Prevention of Relevant Hazards (Cu.G.Ri., Consorzio inter-Universitario per la previsione e la prevenzione dei Grandi RIschi), Via Giovanni Paolo II, 84084 - Fisciano (SA), Italy.

\* Corresponding author: J.S.: sameh.jallouli92@gmail.com; V.N. vnaddeo@unisa.it.

# Highlights

- Remediation of tannery wastewater by sequential electrocoagulation and UV photolysis
- High reduction of chemical oxygen demand COD
- The treatment was time-course monitored by UV-Vis spectroscopy
- Operational costs were evaluated
- Remediated wastewater presents reduced phytotoxicity on *Hordeum Vulgare*

# Graphical Abstract

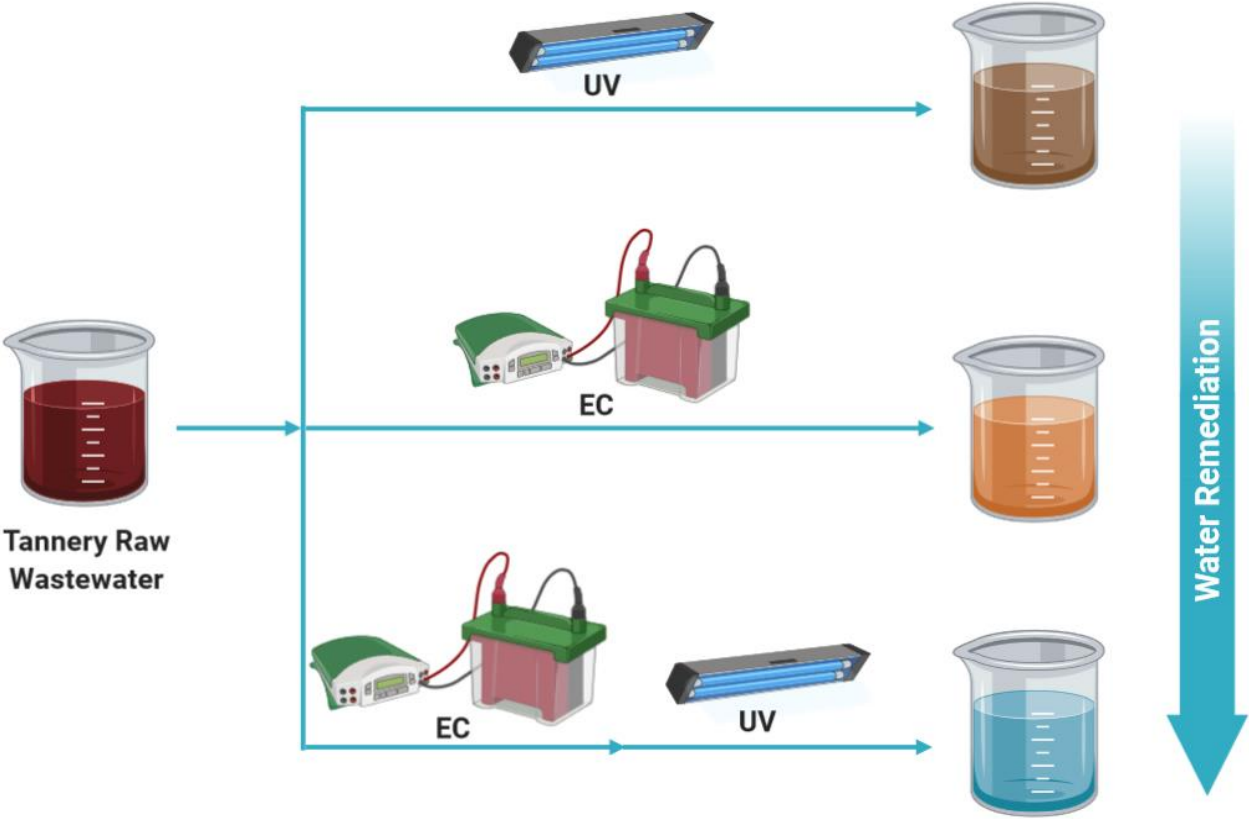

**Abstract**

Tannery wastewater contains large amounts of pollutants that, if directly discharged into ecosystems, can generate an environmental hazard. The present investigation has focused the attention to the remediation of wastewater originated from tanned leather in Tunisia. The analysis revealed wastewater with a high level of chemical oxygen demand (COD) of 7376 mg$_{O2}$/L. The performance in reduction of COD, via electrocoagulation (EC) or UV photolysis or, finally, operating electrocoagulation and photolysis in sequence was examined. The effect of voltage and reaction time on COD reduction, as well as the phytotoxicity were determined. Treated effluents were analysed by UV spectroscopy, extracting the organic components with solvents differing in polarity. A sequential EC and UV treatment of the tannery wastewater has been proven effective in the reduction of COD. These treatments combined afforded 94.1 % of COD reduction, whereas the single EC and UV treatments afforded respectively 85.7 and 55.9 %. The final COD value of 428.7 mg/L was found largely below the limit of 1000 mg/L for admission of wastewater in public sewerage network. Germination tests of *Hordeum Vulgare* seeds indicated reduced toxicity for the remediated water. Energy consumptions of 33.33 kWh/m$^3$ and 314.28 kWh/m$^3$ were determined for the EC process and for the same followed by UV treatment. Both those technologies are yet available and ready for scale-up.

# 1. Introduction

Tannery wastewater is highly polluted in terms of: *i)* suspended solids, *ii)* organic matter (causing high COD and BOD values), *iii)* inorganic compounds such as chlorides, sulphides, nitrogen-containing compounds (ammonia, nitrites and nitrates) [1–3] and heavy metals complexes, in particular of chromium (used as tanning agent) [4]. Sludges derived from tanning processes pose serious hazards to the environment due to their contamination with different agents, pigments, dyes, retaining agents and heavy metals, such as arsenic, cadmium, copper, cobalt, chromium, iron, nickel and zinc [5]. Often in developing countries, those effluents are directly discharged without treatment into rivers, lakes and marine ecosystems [6,7].

The low biodegradability of some tannery pollutants and the relative large occurrence constitutes a global environmental issue [8]. Sutton et al. [9] and Rezgama et al. [10], and relate co-workers, have shown that high levels of nitrogen-containing compounds originate eutrophic conditions, with consequent alteration of the aquatic biomass levels, limitation of biodiversity and acute chronic effects at a specific stage of life. That is also accompanied by an increase of salinity, osmotic potential, total suspended solids and COD. A high value of COD can lead to toxic conditions of water media, due to the persistence of biologically recalcitrant organic substances [11]. Hence, it is indispensable to decontaminate water effluents of tanning industry before discharging into the environment. Water remediation treatments include: pollutant adsorption [12,13], chemical coagulation [14,15], ozonation [16–18], electrooxidation [19,20], ultrasonication [21], Fenton oxidation [22–24], photodegradation [21,24,25] and biological treatments [26,27]. Some of these processes are often expensive, suffer from operational difficulties, and result poorly sustainable because involving the use of large amounts of additional chemical agents. Biological remediation often cannot be directly applied to such polluted typology of wastewater that can cause the depletion of the microorganisms involved in the treatment. Previous researches reported that biodegradation [28], chemical-coagulation [29] and membrane filtration [30] can singularly convert tannery wastewater pollutants to less harmful forms that, however, still not satisfying the environmental regulation in matter of wastewater discharge. Therefore, the development of efficient and cost-effective technologies for the treatment of tannery wastewater is indispensable.

Nowadays, electrocoagulation (EC) treatment of wastewater is gaining a wide interest [31,32]. The EC process consists of producing *in situ* $Al^{3+}$ ions by electrochemical dissolution of an aluminium electrode under application of an electrical potential. At the anode can occur both water and aluminium oxidations, but the oxidation of the latter exceeds that of water due to the electropositivity of this metal (see Scheme 1a). On the other electrode, the cathode, occurs the reduction of the water to gaseous hydrogen (Scheme 1b). The evolution of gaseous hydrogen plays an additional beneficial role to the flotation and coagulation of pollutants. The overall reaction is reported in Scheme 1c.

EC generated aluminium cations, in addition to the coagulation of insoluble salts (e.g. phosphates and sulphates), afford the precipitation of organic and inorganic colloids, with resulting abatement of total suspended solids including those containing heavy metals. EC processes afford the removal even of the smallest colloidal particles. The sludges formed during EC treatments tend to be readily settable, facilitating the subsequent removal from the treated water. EC avoids excessive use of coagulating agents for water remediation, preventing secondary pollutions caused by the addition of these chemicals.

Jotin et al. [33] for the treatment of landfill leachates, with an initial pH equal to 8 and COD value of 16,464 mg/L, found that the optimal conditions were a current tension of 12 V and an electrolysis time of 100 min, with resulting 62.7% of COD reduction. Ameziane et al. [34] treating hospital wastewater by EC achieved a reduction efficiency of 79.2 % for COD, 93.7% for suspended solids and 97.3 % for the total faecal coliforms. Farhadi et al. [35] studied the COD reduction in pharmaceutical wastewater adopting iron electrodes, obtaining 34.2% of COD reduction with an electric energy consumption of 65.06 kWh/kg$_{COD}$.

The combination of electrochemical treatments with other treatments, *i.e.*: advanced oxidation processes [36], UVC/UVB photolysis [37] in presence of $TiO_2$ [38] or $H_2O_2$ [39] have gained an increasing interest in the community of researcher involved in the field of water remediation. Currently, UV photolysis is extensively applied to wastewater treatment, particularly in disinfection treatments. Therefore, the technology suitable for the extension of this process to other modalities of water remediation is yet available. In this process, the energetic radiation strongly affects the organic component of wastewater causing cleavage of chemical bonds, rearrangements, as well as redox reactions (see Scheme 2). UV radiation can induce degradation of inorganic and in particular organic molecules by direct photolysis via three main modalities: homolytic and heterolytic cleavage of molecules or via photoionization [40].

Both EC and UV treatments are easy to operate and require simple equipment with reduced maintenance needs. Recent studies from Alvarez et al. [41] demonstrated that the application of electrochemical and ultraviolet (EC-UV) treatments allowed to achieve high effluent decolourization, up to 99%, and the decomposition of small amounts of haloforms generated during the electrochemical treatment.

a) Anode:

$$Al_{(s)} + 3\,OH^- \longrightarrow Al(OH)_{3(aq)} + 3\,e^- \qquad (E^0_{red} = -2.31\ V)$$

$$2\,H_2O \xcancel{\longrightarrow} O_{2(g)} + 4\,H^+ + 4\,e^- \qquad (E^0_{red} = +1.229\ V)$$

b) Cathode:

$$2\,H_2O + 2\,e^- \longrightarrow H_{2(g)} + 2\,OH^- \qquad (E^0_{red} = -0.827\ V)$$

c) Overall redox reaction:

$$2\,Al_{(s)} + 6\,H_2O \longrightarrow 2\,Al(OH)_3 + 3\,H_{2(g)} \qquad (E^0_{Al^{3+}/Al} = -1.662)$$

**Scheme 1.** Redox reactions in EC processes.

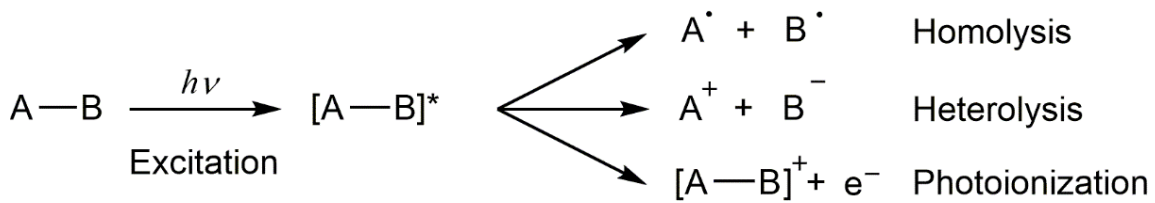

$$A\!-\!B \xrightarrow[\text{Excitation}]{h\nu} [A\!-\!B]^* \begin{cases} A^{\bullet} + B^{\bullet} & \text{Homolysis} \\ A^+ + B^- & \text{Heterolysis} \\ [A\!-\!B]^{+} + e^- & \text{Photoionization} \end{cases}$$

**Scheme 2.** Mechanism of UV decomposition of chemical compounds.

In the current study, a sequential EC-UV treatment of wastewater was applied as an alternative method for the remediation of tannery effluent. The effect of different voltages on EC process by measuring the COD removal efficiencies was initially investigated. EC-UV method is proposed as a viable alternative approach for the treatment of tannery wastewater, as a part of a sustainable and integrated wastewater management. The appropriateness of treated tannery wastewater effluent, following the criteria for water recycling and reuse in Mediterranean countries, was investigated.

The performance of various technologies for the treatment of tannery effluents is summarized in Table 1. The adoption of coagulation and flocculation using *Musa* sp. flower extract resulted effective in the removal of recalcitrant compounds that are not easily removed by other methods [42]. Besides, the treatment improved the biodegradability of the effluent. However, coagulation and flocculation processes do not afford the decomposition of the organic pollutants that are simply removed from the wastewater. Electrocoagulation of pollutants in tannery wastewater using aluminium electrodes was also investigated [43]. A COD removal of 68% was obtained after 60 minutes. However, any evaluation of the potential toxicity of the final effluents after treatment was assessed. The feasibility of photoassisted-electrochemical oxidation (PEO) of tannery effluents was also proposed [44]. A significant COD removal of 92% was achieved with the PEO process. Compared with other described processes, the reaction time of 18 h resulted quite limiting. Moradi et al. [37] studied the electrocoagulation process combined with UVC/VUV photoreactor to degrade tannery wastewater with a COD loading of 11.666 mg/L. A COD removal efficiency of 99.52% was achieved after 60 minutes of treatment. The authors stated that the selected combination is an efficient and promising method for the treatment of real industrial tannery wastewater.

In this context, our study demonstrates the possibility of treatment of real tannery wastewater with COD removal efficiency reaching 94.1% after 8 h of EC followed by UV photolysis. The low final COD value of 428.7 mg/L result largely below the limit for admission of wastewater in public sewerage network of Tunisia (see Table 2) [45]. The knowledge in this contest will be helpful for developing an alternative and effective treatment method of tannery effluent and other similar types of wastewaters.

**Table 1:** Comparison of different remediation methods of tannery wastewater.

| Process | Treatment time | Reactor volume (mL) | Significant findings | Advantages | References |
|---|---|---|---|---|---|
| Coagulation and flocculation using Musa sp. flower extract | 1st step: 3 min, 2nd step: 12 min, 3rd step: 15 min. | 2000 | Turbidity reduction: 78%; Total chromium removal: 65.4%; Hexavalent chromium removal: 39.43%; Trivalent chromium removal: 61.02%; Biodegradability: satisfactory. | Designed to remove recalcitrant compounds that are not easily removed by other methods. | Pinto et al. [42] |
| Electrocoagulation | 60 min | 2000 | COD reduction of 68% by mild steel and aluminium electrodes. | Characterized by simple equipment; easy operation; no chemical addictive and a lesser amount of precipitate or sludges. | Feng et al. [43] |
| Photoassisted-electrochemical oxidation (PEO) | 18 h | 400 | COD reduction: 92%. | Production of radicals to degrade the COD effectively without any sludge. | Selvaraj et al. [44] |
| Electrocoagulation combined to UVC/VUV | 60 min | 180 | COD reduction: 99.52%; Total chromium removal: 100%; Sulphide removal: 98.27%. | The selected combination is an efficient and promising method for the treatment of real industrial tannery wastewaters. | Moradi et al. [37] |
| Sequential Electrocoagulation/UV | EC: 5 h; UV: 3 h. | 300 | COD reduction: 94.1%; Low final COD value: 428.7 mg/L (largely below the limit for admission of wastewater in public sewerage network). | The simplicity of operation. No chemicals are needed. There is no possibility of secondary pollution due to high concentrations of chemicals for remediation. High COD removal efficiency. | Current study |

## 2. Experimental section

### 2.1. Materials

The tannery wastewater was collected from Textiles & Leather Company in Moknine, located in the centre of the eastern Sahel in Tunisia. The organic solvents used for the extraction of organic compounds from leather industry wastewater, namely 1-butanol (99 %) and ethyl acetate (99.8 %), were purchased from Sigma-Aldrich (Germany) and, unless otherwise stated, used as received without any further purification procedure. Standard chemical reagents for COD determination were purchased from Sigma-Aldrich (Germany).

### 2.2. Instrumentation and methods

Physicochemical parameters were determined using standard analytical procedures. Total suspended solids (TSS) and total volatile solids (TVS) were determined according to APHA methods [46], determining weight loss after drying the sample overnight at 105°C and treatment for 2 h at 600°C. Clarified effluents were obtained by removal of eventual debrides and coarse particles from raw effluents by filtration with filter paper (average pore size of 0.63 µm). Sulphide concentrations were determined volumetrically according to the procedure reported by Bouzid et al. [17]. Chloride concentrations were determined according to reported standard methods [47]. Conductivity and pH of effluents were respectively measured using an AD332 conductometer and an inLab7110 pH meter. UV absorption spectra were acquired by using a Jenway 7315 spectrophotometer. Ammonium concentrations were determined using a Buchi K360 automated Kjeldahl apparatus. Chemical oxygen demand (COD) was determined according to the procedure reported by Knechtel et al. [48]. COD reduction efficiencies were evaluated with the following equation:

$$\textbf{COD reduction efficiency (\%)} = \frac{COD_i - COD_f}{COD_i} \times 100 \qquad \textbf{Eq.1}$$

where $COD_i$ and $COD_f$ were the initial and final value of COD, respectively before and after the treatment.

### 2.3. Liquid-liquid extraction

Pollutants were separated from wastewater by liquid-liquid extraction according to the method of Natarajan et al. [49], using organic solvents with different polarity. 250 mL of wastewater were washed with 50 mL of 1-butanol or ethyl acetate for three times. The organic phases were collected and concentrated with a constant stream of nitrogen to the final volumes of 1 mL. The residual oil was dissolved in ethyl acetate or 1-butanol and analysed by UV-Vis spectroscopy.

## 2.4. Electrocoagulation process of tannery wastewater

The electrocoagulation was carried out at 25°C ±3°C in batch mode in a 1 L glass reactor using rectangular plate aluminium electrodes with size of 75×40 mm and inter electrodes distance of 30 mm. A TM 501-2 DC power supply was used at current voltages of 5, 12 and 18 V, and current density of 12 A/cm .

## 2.5. Photochemical UV treatments of tannery wastewater

UV photolysis was conducted in a 120 mL cylindrical laboratory-scale photoreactor, operating in a closed recirculating circuit driven by a centrifugal pump. Tannery effluents were irradiated with a low-pressure mercury lamp (Philips, Holland) with power of 11W and wavelength range of 200–280 nm

## 2.6. Phytotoxicity assay of wastewater

The phytotoxicity assays were conducted using seed germination and growth parameters of *Hordeum Vulgare* seeded in raw wastewater, treated wastewater, saltwater or deionized water in Petri dishes incubated at 23±1 °C, as described by Kumari et al. [50]. Seedling growth parameters, root and shoot length, were evaluated after 5 days of treatment and compared to control experiments where *Hordeum Vulgare* was seeded in distilled water.

# 3. Result and discussion

## 3.1. Characteristic of the untreated tannery effluent

A comparison of the physicochemical parameters of the tannery wastewater determined in the current and, as references, for other studies reported in literature, is presented in Table 1. Table 2 reports the limit values for these parameters for admission of the wastewater in public sanitation network of Tunisia [45], Italy (the Italian values are representative the rest of Europe) [51] and USA [52].

The investigated wastewater resulted highly alkaline with a pH of 10.78. High value of COD (7376 mg/L), TSS (2400 mg/L), sulphides (204 mg/L), ammonium (193.5 mg/L) chlorides (1573 mg/L) and conductivity (91.9 mS.cm$^{-1}$) were also determined. The determined parameters exceed the permissible limit values for admission of wastewater in Tunisian public sanitation network (D.G. n° 2018-315) [45]. The direct discharge of such effluent in aquatic ecosystems can severely affect sunlight penetration and aquatic photosynthetic processes, with a strong impact on flora and fauna exposed, as well as, in turn poses serious hazards on the human health [53,54]. The high value of COD found of 7376 mg/L can be ascribed to the presence of oxidizable organic matter and nitrogen-containing compounds, released during the treatment of the leather, and to oxidizable salts added for the tanning. The presence of high content of sulphides and sulphates is associated respectively to the use of sodium sulphide and sulfuric acid in leather processing [55]. Galiana-Aleixandre et al. [56] reported that high concentrations of sulphide (204 mg/L) in wastewater cause serious environmental problems of odour, as well as anaerobic and corrosive conditions. Finally, the concentration of chromium resulted slightly greater than that acceptable for the discharge of the wastewater (Table 1).

**Table 2.** Physicochemical characteristics of wastewater investigated in the current study in comparison with those reported in literature and the permitted limit values in Tunisia, Italy and USA, as a reference.

|  |  | pH | Conductivity (mS/cm) | COD (mg/L) | TSS (mg/L) | TVS (mg/L) | [Cl$^-$] (mg/L) | [S$^{2-}$] (mg/L) | [NH$_4^+$] (mg/L) | [Cr$^{6+}$] (µg/L) |
|---|---|---|---|---|---|---|---|---|---|---|
| **Limit values for wastewater admission in public treatment network** | Tunisia (D.G. n° 2018-315)[45] | 6.5-9.0 | 5 | 1000 | 400 | - | 700 | 3 | 50 | 500 |
|  | Italy (Europe) (D.Lgs n. 152) [51] | 5.5-9.5 | - | 500 | 200 | - | 1200 | 2 | 30 | 200 |
|  | USA [52] | 5-11 | - | 1500* | 900 | - | - | - | 200 | 6000 |
| **Pinto et al. [42]** |  | 10.66 | - | 3229 | - | - | - | - | - | - |
| **Isarain-Chávez et al. [57]** |  | 4.3 | <6.3 | <9922 | <530 | - | 1239 | <28.9 | - | <0.11 |
| **Feng et al. [43]** |  | 7-8.5 | 8-10 | 8-10 | - | - | 2700-2800 | 100-120 | - | - |
| **Ben Amar et al. [58]** |  | 11.8 | 3.5 | 4500 | - | - | - | 102 | - | - |
| **Current study** |  | 10.78 | 91.9 | 7375.7 | 2400 | 82 | 1573 | 204 | 193.5 | 554 |

**Table 3.** Organic compound pollutants identified in tannery effluents extracted by different solvents

| Extraction solvent | Organic compounds extracted | References |
|---|---|---|
| **Acetonitrile + acetone, methanol or dichloromethane** | 2,4-bis(1,1-dimethyl)phenol; 10-methylnonadecane; docosane; bis(2-ethylhexyl)phthalate; 2,6,10-dodecatrien-1-ol-3,7,11-trimethylacetate; 1,3-hexadien-5-yne; 1,2-benzenedicarboxylic acid, diisooctyl ester; 2,2,3-Trimethyl oxepane; benzene; 3-nitrophthalic acid | Zubair Alam et al. [59] |
| **Chloroform** | 2-phenylethanol; nonadec-1-ene; (3R,6S)-3-Methyl1-6(-prop-1en-2-yl)-cyclohex 1-enolbis(2-methoxyethyl) phthalate; hexatriacontane; heneicosane; 2,4-di-tert-butylphenol tricosane; 2,3-epoxypinane | Natarjen et al. [49] |
| **Ethyl acetate** | Benzene acetic acid; benzoic acid; benzene propanoic acid; 2-methylbutanoic acid; propanoic acid | Chandra et al. [60] |

## 3.2. Electrochemical reduction of COD: effect of voltage and treatment time

Voltage and current density are critical parameters in EC processes for pollutant removal in wastewater. Current density strongly affects the rate of metal hydroxides formation at the electrode interface and, in turn, the overall efficiency of the EC process. Previous works have shown that the increase of voltage in EC leads to a quickest treatment [61–63]. On the other hand, the increase of the EC voltage produces a higher oxidation of the aluminium electrode, with resulting reduced electrode lifetime and greater production of sludges.

The effect of the applied voltage on the COD reduction efficiency was examined and showed in Figure 1. EC experiments were carried out for 90 min for each run by varying the voltage (5V, 12V, and 18V). Wastewater aliquots were sampled at intervals of 10 min in order to follow the course of the reaction. Our attention was dedicated to the abatement of the COD value, as an indicator of the contamination level of the wastewater. Electrodes were substituted after each run with the purpose of reduce the effects of corroded electrodes on the determination of COD reduction efficiency. It is appreciable that at the voltage of 5V the efficiency in COD reduction constantly increased with the time of treatment, reaching the value of 58% after 90 min (red histogram bars in Figure 1). This result is in agreement with the data previously reported by Wang et al. [64] for COD reduction in laundry wastewater by electrocoagulation/electroflotation by applying the same voltage. By increasing to 12 V the electric potential, the efficiency rapidly increased within the first 30 minutes, subsequently reaching a plateau value at prolonged reaction time. On the contrary, at the voltage of 18 V the COD reduction efficiency increased slowly during the first 30 minutes, reaching high performances only at prolonged treatment time. This behaviour can be ascribed to the formation of compounds, refractory to further electrochemical oxidation on the surface of electrode and to the rapid passivation of the electrode at the high voltage applied. The reduced formation rate of aluminium ions inhibits the removal of pollutants. A prolonged treatment time restores the surface of the electrodes and increases the release rate of aluminium ions, allowing high values of COD reduction. The EC performance, both in terms of pollutants removal and energy consumption, is strongly influenced by the electrolysis time. The effect of a prolonged treatment time was explored by applying a voltage of 5V (Figure 2). After 5 h of treatment, 85% of COD reduction was achieved. This result is remarkable when compared to the performance previously reported in literature. An efficiency in COD reduction of 90.2 % of in 12 h and of 25% in 2.5 h were respectively reported by Keerthi et al. [65] and by Ouaissa et al. [66].

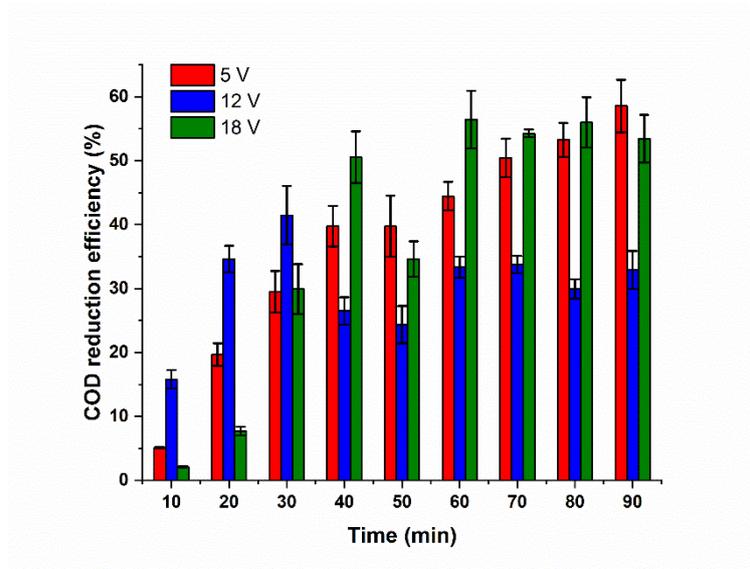

**Figure 1.** COD reduction efficiency at variance of voltage and electrolysis time.

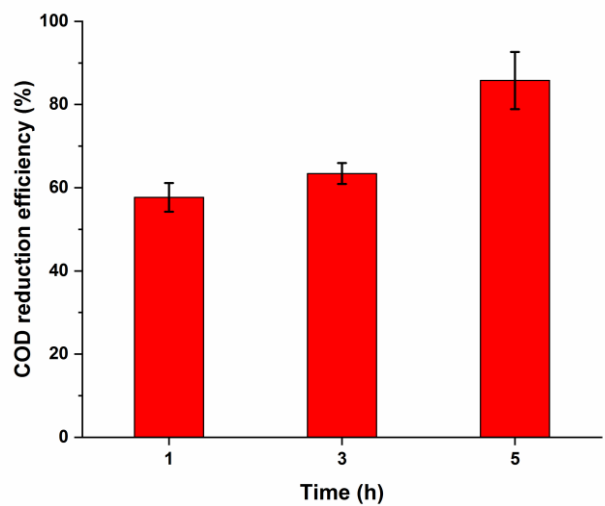

**Figure 2.** COD reduction as a function of time using EC at the voltage of 5V.

### *3.3. Time-course investigation on EC treatment of tannery wastewater by UV-Vis spectroscopy*

Tannery wastewater is a quite complex chemical mixture and the corresponding analysis by UV-Vis spectroscopy can provide a rapid depiction of the level of contamination with organic and inorganic compounds responsive in the UV-Vis spectral width.

Figure 3 shows the UV-Vis absorbance changes of tannery wastewater during the EC treatment at the applied voltage of 5V. Two broad adsorption bands were found centred at wavelength of 255 and 275 nm. The first band can be principally ascribed to absorption due to π-π* transitions in isolate or conjugated C=C bonds, *e.g*. in organic aromatic compounds. The second band centred at wavelength of 275 nm can be assigned to *n*-π* transitions in carbonyl groups. In addition, in the latter spectral width are located absorptions due to ligand-to-metal charge transfer (LMCT) in inorganic metal complexes. Therefore, UV-Vis is a powerful tool for investigation on the removal of organic and inorganic pollutants. Figure 3 shows that both the above-mentioned adsorption bands were attenuated by the progression of the electrocoagulation process. Feng et al. [43] reported that the UV-Vis absorbance of wastewater decreased when the EC treatment was applied. Aluminium ions released at the anode (see Scheme 1) act as a coagulant for water-soluble or suspended pollutants. UV-Vis spectroscopy also provided information on the effective decolouration of the wastewater during the EC-UV treatment.

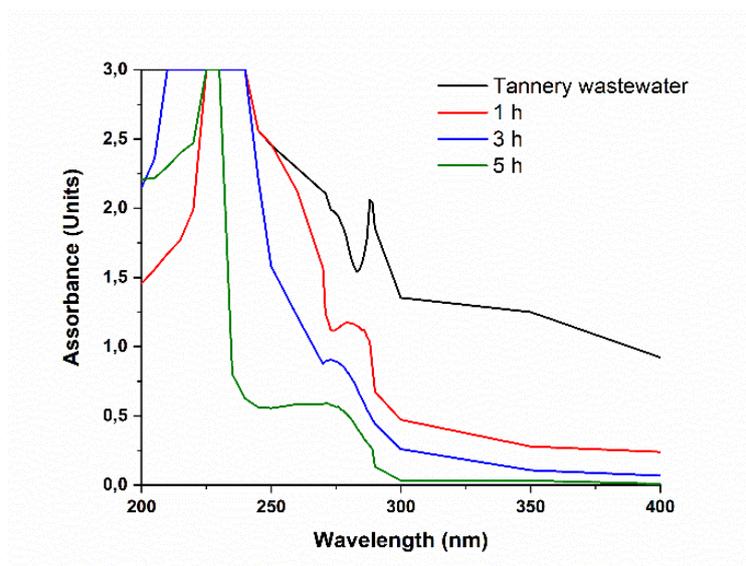

**Figure 3.** UV-Vis spectra normalized of tannery wastewater before and during EC treatment (initial COD value: 7375.7 mg/L; applied voltage: 5 V; current density: 12 A/cm ).

The acquisitions of UV absorption spectra of chemical compounds present in tannery effluents before and after EC treatment were carried out after liquid-liquid extractions (see Figure 4 and 5). The liquid-liquid extractions were carried out by using ethyl acetate and 1-butanol. These solvents have different polarity with dielectric constant of 6 and 18 respectively [67,68]. Ethyl acetate and 1-butanol afford respectively the extraction of chemical compounds with lower and higher polarity, according to their dielectric constants. The compounds extracted by ethyl acetate, both from raw and treated tannery wastewater, show an UV-Vis absorption band centred at wavelength of 255 nm, while those extracted with 1-butanol present a prevalent absorption band centred at 275 nm (compare Figure 4 and 5). On the basis of the solvent polarity and the above-indicated considerations on the UV-Vis absorption bands, can be assumed that ethyl acetate and 1-butanol respectively extract hydrophobic and hydrophilic compounds. Previous investigations on the identification of compounds isolated by liquid-liquid extractions with several solvents support this assumption (see Table 2).

Hydrophilic compounds, with prevalent absorption band at 275 nm (Figure 4), resulted easily removed during the electrochemical process (Figure 4). As matter of facts, Ma et al. [69] showed that such compounds were degraded into low molecular weight derivatives, easily removed by electrocoagulation. In contrast, hydrophobic compounds, and in particular the aromatic ones, result scarcely removed (Figure 5).

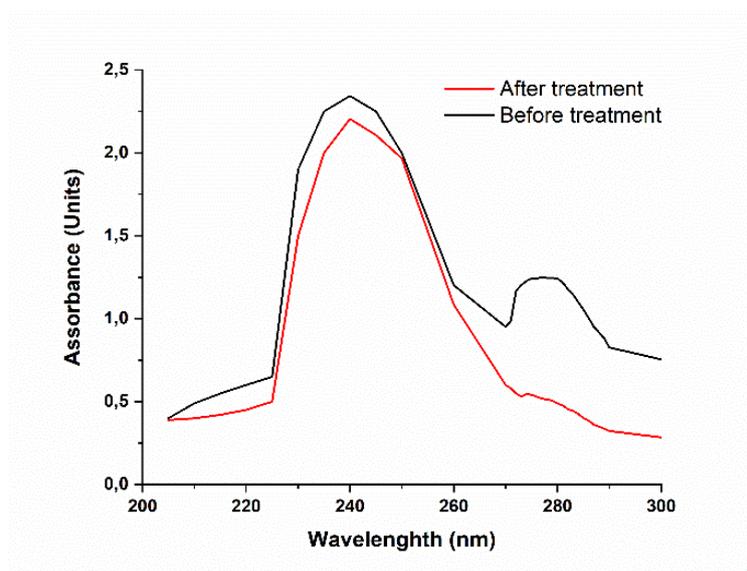

**Figure 4.** UV-Vis spectra of extraction products with 1-butanol from tannery effluent before and after EC treatment (initial COD value: 7375.68 mg/L; voltage: 5V; current density: 12 A/cm ; electrolysis time 5h).

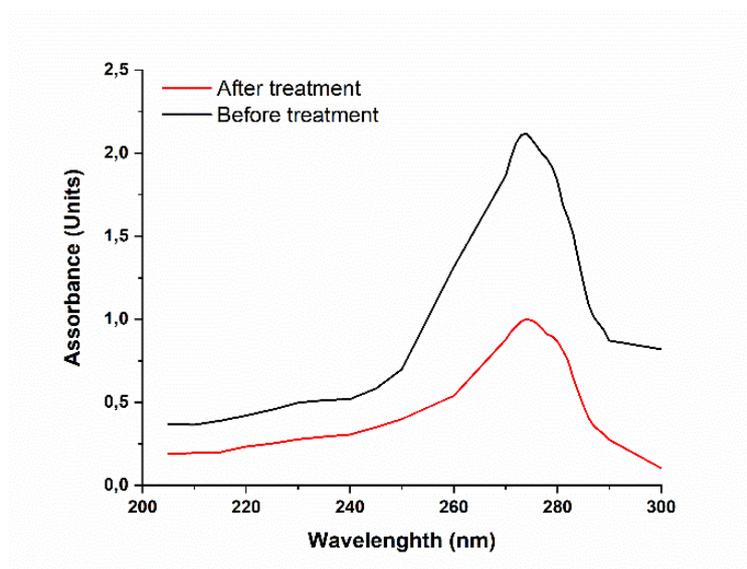

**Figure 5.** UV-Vis spectra of extraction products with ethyl acetate from tannery wastewater before and after EC treatment (initial COD value: 7375.68 mg/L; applied voltage: 5V; current density: 12 A/cm ; electrolysis time 5h).

*3.4. Treatment of tannery effluent by electrocoagulation and UV photolysis*

The COD reduction after sequential EC and UV treatment in tannery wastewater was determined and compared with the corresponding single treatments. The main results are reported in Table 3 and depicted in Figure 6. The single EC and UV treatments respectively afforded a COD reduction of 85.7 and 55.9 %. Interestingly, a sequential process consisting of 5 h of EC treatment followed by 1 h of UV exposure allowed a COD reduction of 94.1 % (Table 3). When an EC process is applied, the persistence of recalcitrant compounds occurs (compare Figure 6 and 3). In order, to overcome this problem, UV photolysis can be applied (see Table 3 and Figure 6).

The final low concentration of COD of 428.74 mg. $L^{-1}$ obtained by the EC-UV treatment satisfies the Tunisian standard for discharge of wastewater (1000 mg/L of COD, see Table 1) [45]. Besides, at the end of the remediation process was observed a reduction of the chromium concentration from 554 µg/L to 430 µg/L, that can be mainly ascribable to the effect of the EC stage. Also in this case, the value resulted below that for admission of the effluent in Tunisian public sanitation network (see Table 2).

**Table 4.** COD values and COD reduction efficiencies after EC, UV and EC-UV.

| Treatment | Treatment time (h) | COD (mg/L) | COD reduction efficiency (%) |
|---|---|---|---|
| EC | 5 | 1018.1 | 85.7 |
| UV | 3 | 3252.0 | 55.9 |
| EC+UV | 5+3 | 428.7 | 94.1 |

Initial COD value of raw wastewater: 7375.68 mg/L. EC reaction conditions: applied voltage: 5V; current density: 12 A/cm .

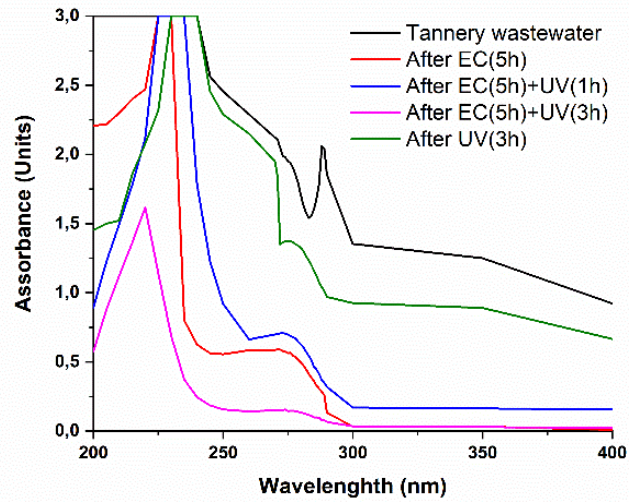

**Figure 6.** UV-Vis spectra of tannery wastewater before and after EC, UV and sequential EC-UV treatment. (Voltage = 5V, current density = 12 A/cm )

## 3.5. Energy consumption

The energetic consumption has a deep impact on the analysis of the costs of industrial wastewater treatment. Energy consumption normalized with respect to the volume of wastewater treated, for electrocoagulation processes ($E_{EC}$), can be determined with the following equation:

$$E_{EC} = \frac{VIt}{v_R} \qquad \text{Eq. 2}$$

where $V$ is voltage (V), $I$ is applied current (A), $t$ is electrolysis time (h) and $v_R$ is the volume of effluent (L) [70,71].

Energy consumption for the photolysis ($E_{EC}$) was calculated using the following equation:

$$E_{UV} = \frac{P_{el} \times t \times 1000}{\log\frac{COD_i}{COD_f} \times v_R \times 60} \qquad \text{Eq. 3}$$

Where $P_{el}$ is the electrical power (kW), $t$ is the irradiation time (h), $v_R$ is the volume of effluent (L) and $COD_i$ and $COD_f$ are initial and final COD concentrations (mg/L).

The total power consumption of the EC process followed by the UV photolysis is given by the sum of the previous values:

$$\boldsymbol{E_{EC-UV}} = E_{EC} + E_{UV} \qquad \textbf{Eq. 4}$$

In this study, the energy consumptions associated with EC, UV and EC-UV treatment were respectively of about 33.33 kWh/m$^3$, 314.28 kWh/m$^3$ and 347.61 kWh/m$^3$, that well compare to the data reported by Selvaraj et al [44]. The authors demonstrated that after the photoelectrochemical oxidation, the COD removal and current efficiency were equal to 92% and 26.6%, respectively. However, by electrooxidation, only 62% of COD removal was achieved with 18% of current efficiency.

### *3.6. Evaluation of the phytotoxicity of untreated and treated tannery wastewater*

The effect of raw and treated tannery wastewater, as well as of salt water as a reference, on germination of barley (*Hordeum vulgare*) seeds in tests with a duration of 5 days, was investigated by determination of the number of germinated seeds (Figure 7). Specific ionic effects were yield evident carrying out the experiments at iso-molar concentrations. The seed germination in untreated tannery wastewater resulted strongly inhibited when compared to the germination in treated wastewater. Results depicted in Figure 6 show clearly that tannery wastewater was phytotoxic, inhibiting both seed germination, rooting and apical growth of *Hordeum Vulgare* seeds. Similar effects were reported by Yadav et al. [72] by assessing the influence of tannery wastewater on the germination of mung bean (*Vigna radiata*). Besides, when the tests were performed using salt water, the results have shown that salts negatively affect seed germination in a similar manner to the test carried out in treated wastewater. This finding suggests an important role for the salt concentration on the germination of the *Hordeum Vulgare* seeds. Negative effects on germination and symptoms of toxicity were previously observed by Alaoui et al. [73] for high concentrations of salts in water.

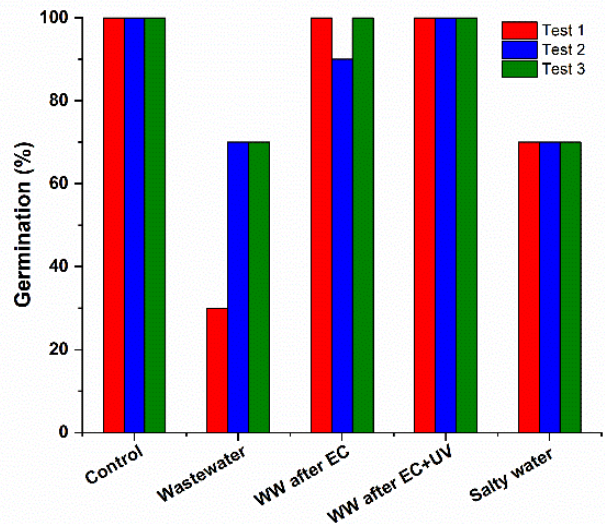

**Figure 7.** Phytotoxicity tests.

## 4. Conclusions

A novel process for the abatement of pollutants in tannery wastewater has been investigated. The process resulted efficient and cost-effective. Voltage and time of electrolysis in the electrocoagulation process of tannery wastewater were investigated. A voltage of 5V was found effective from the point of view both of the performance in wastewater remediation and of the containment of the energetic costs. The course of the EC process has been investigated by UV-Vis spectroscopy by liquid-liquid extractions of organic pollutants from wastewater with ethyl acetate or 1-butanol, identifying possible compounds recalcitrant to the treatment.

A sequential EC and UV treatment of tannery wastewater has been proven effective in the reduction of the COD. Those treatments afforded 94.1 % of COD reduction, whereas the single EC and UV treatments afforded respectively 85.7 and 55.9 %. The final COD value of 428.7 mg/L was found largely below the limit of 1000 mg/L for admission of the wastewater in Tunisian public sewerage network. Germination tests of *Hordeum Vulgare* seeds in raw and treated wastewater indicated reduced toxicity for the remediated water. An energy consumption of 33.33 kWh/m$^3$ and 314.28 kWh/m$^3$ was determined for the EC process and for the same followed by UV treatment.

## Acknowledgement

Sameh Jallouli wishes to thank the components of the laboratory "Laboratoire des Sciences de l'Environnement LARSEN" for geo-chemical analyses, Khawla Chouchene and Dalinda Khoufi for assistance in the revision of the manuscript and Mohamed Seddik Mahmoud Bougi for technical assistance. The authors would like to express gratitude for support from University of Salerno: (FARB grants) and the Inter-university centre for prediction and prevention of relevant hazards (Consorzio inter-Universitario per la previsione e la prevenzione dei Grandi Rischi, C.U.G.RI.).## Declaration of Competing Interest

The authors of this manuscript declare that there is no conflict of interest.

Figure 1

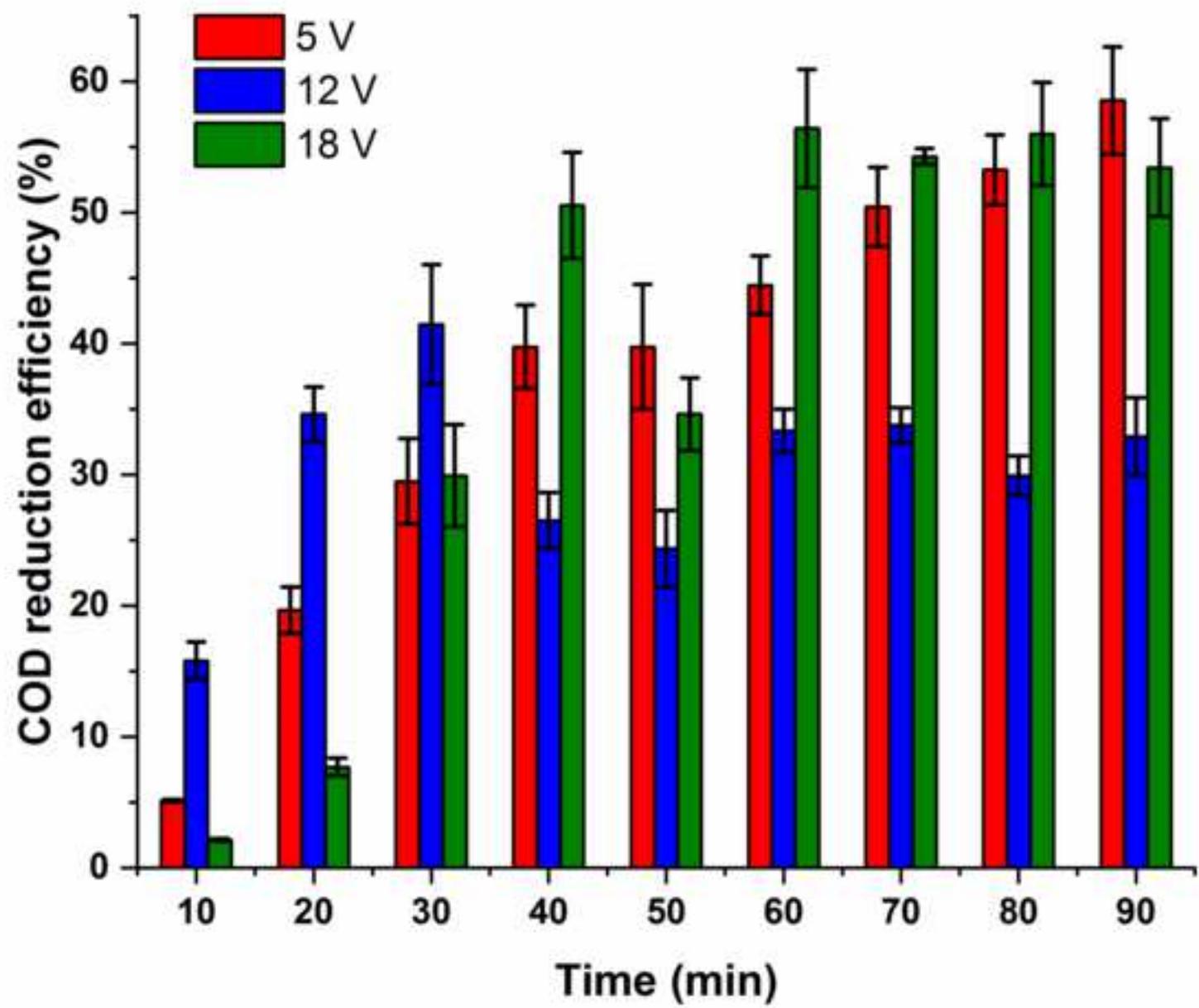

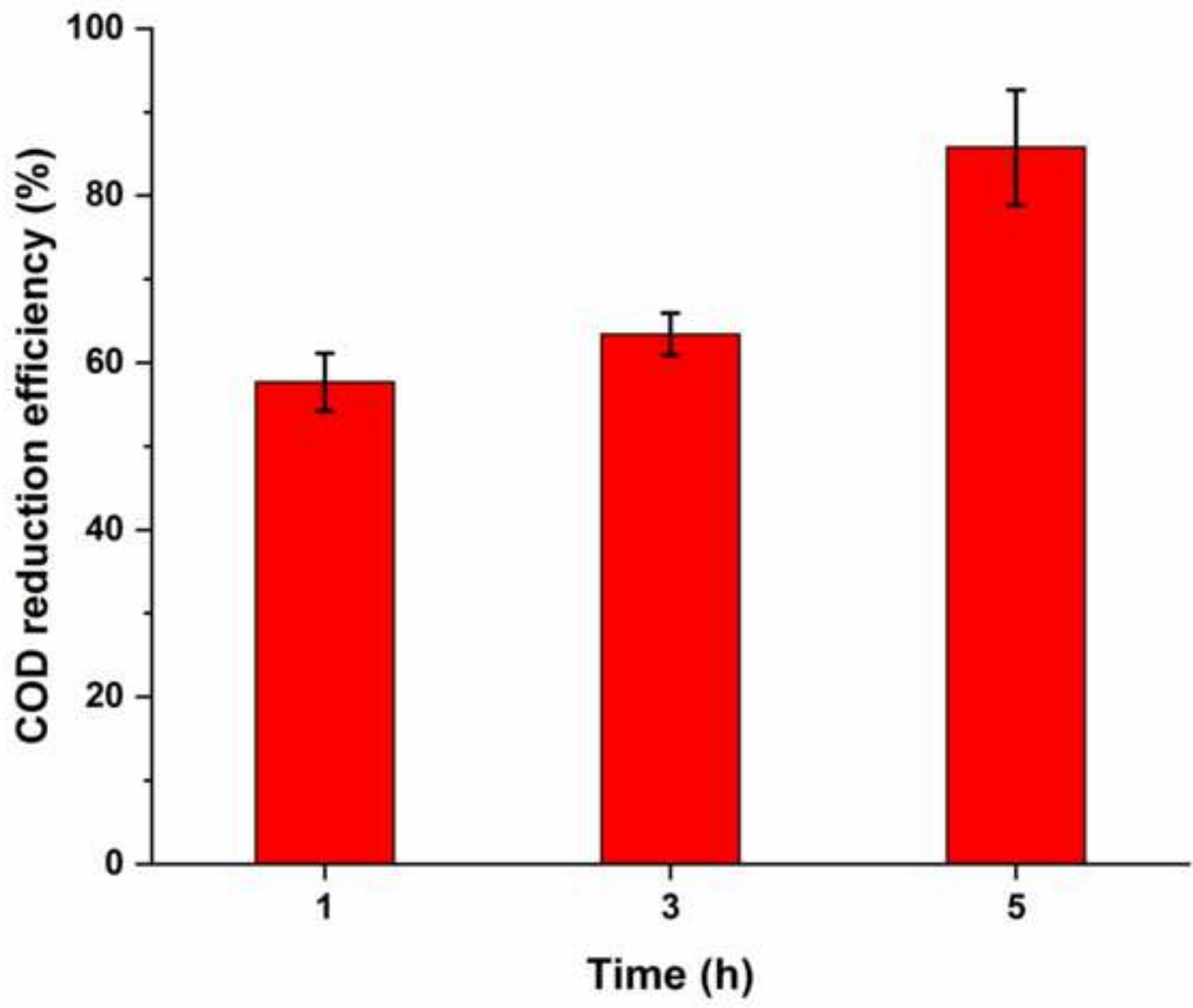

Figure 3

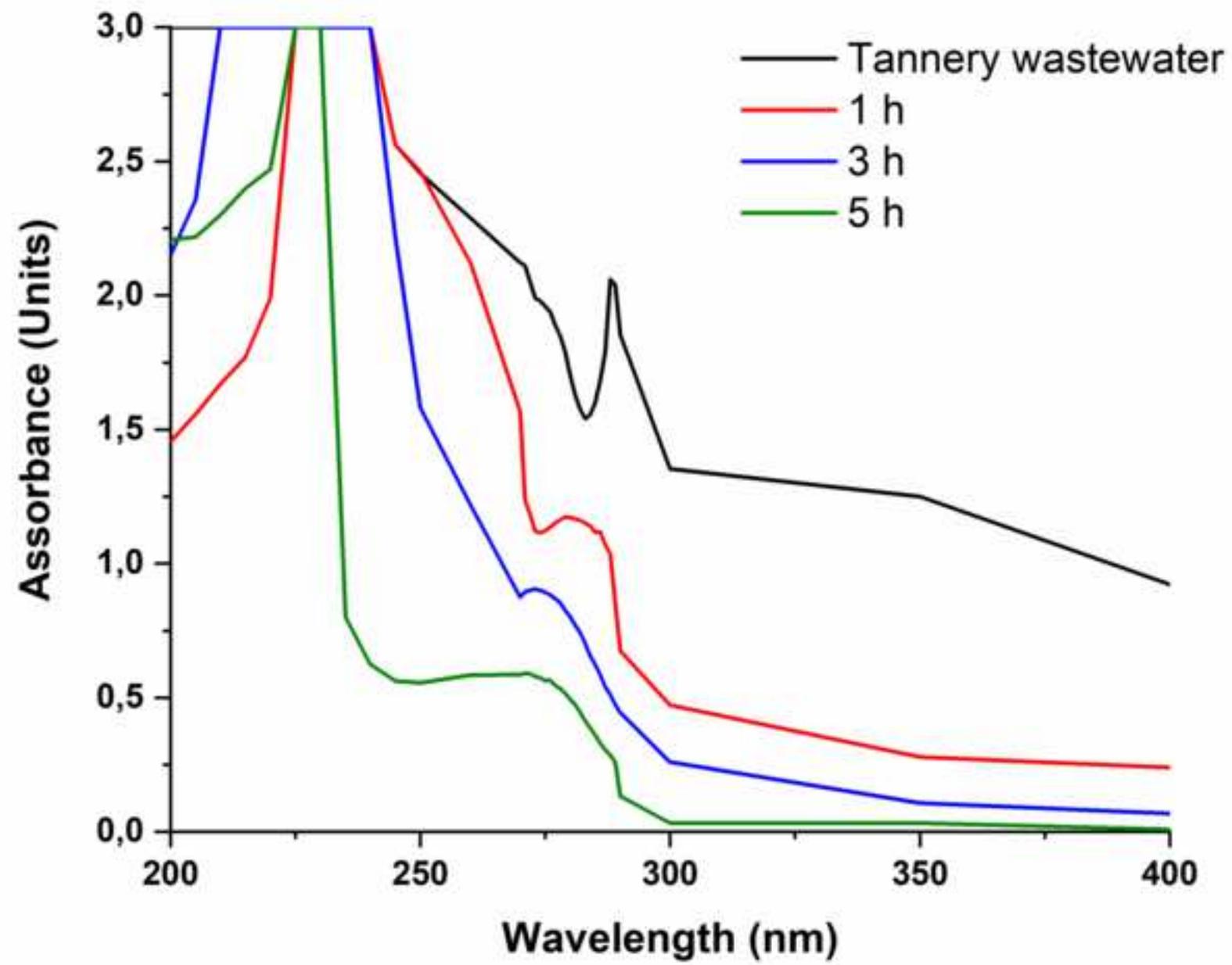

Figure 4

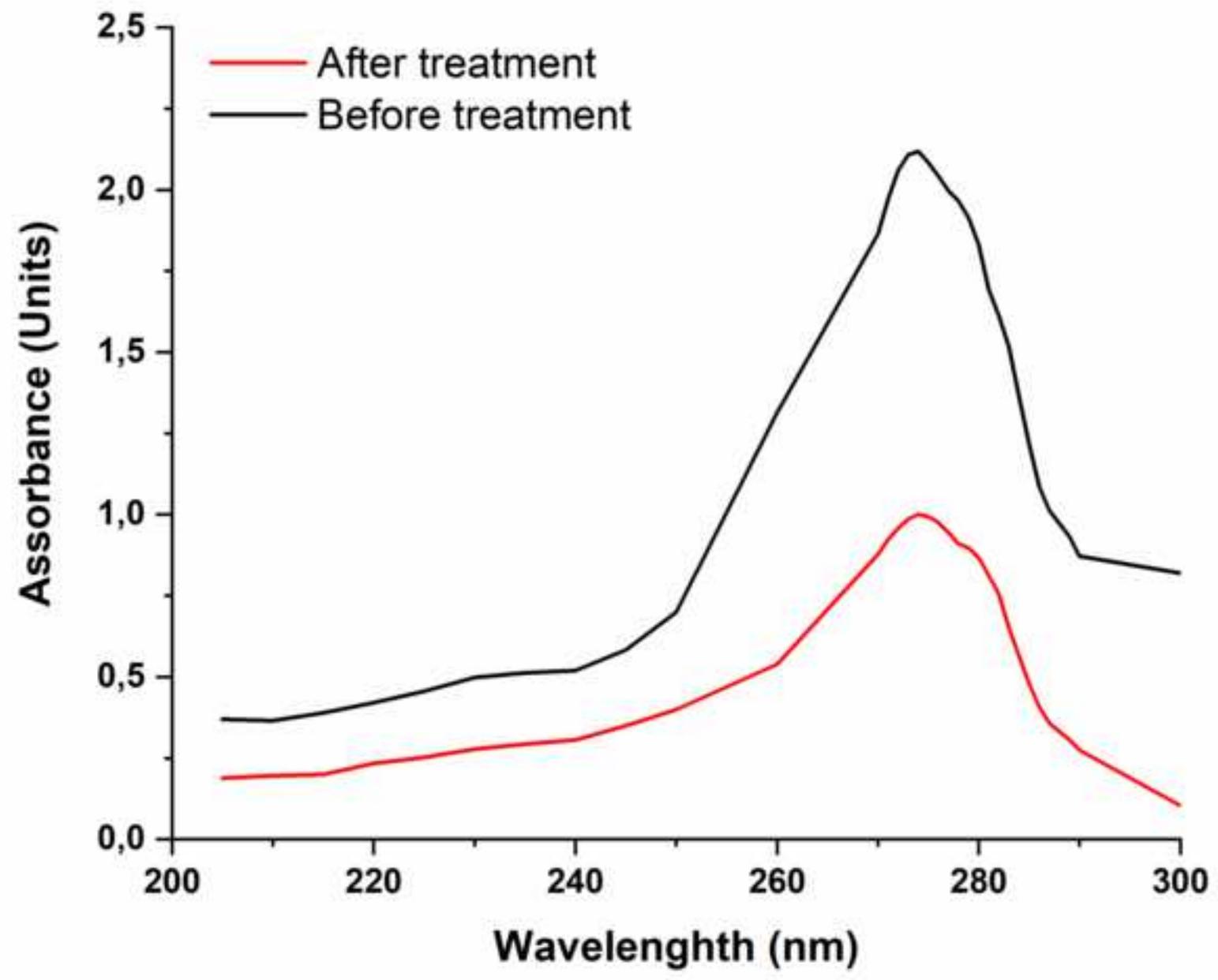

Figure 5

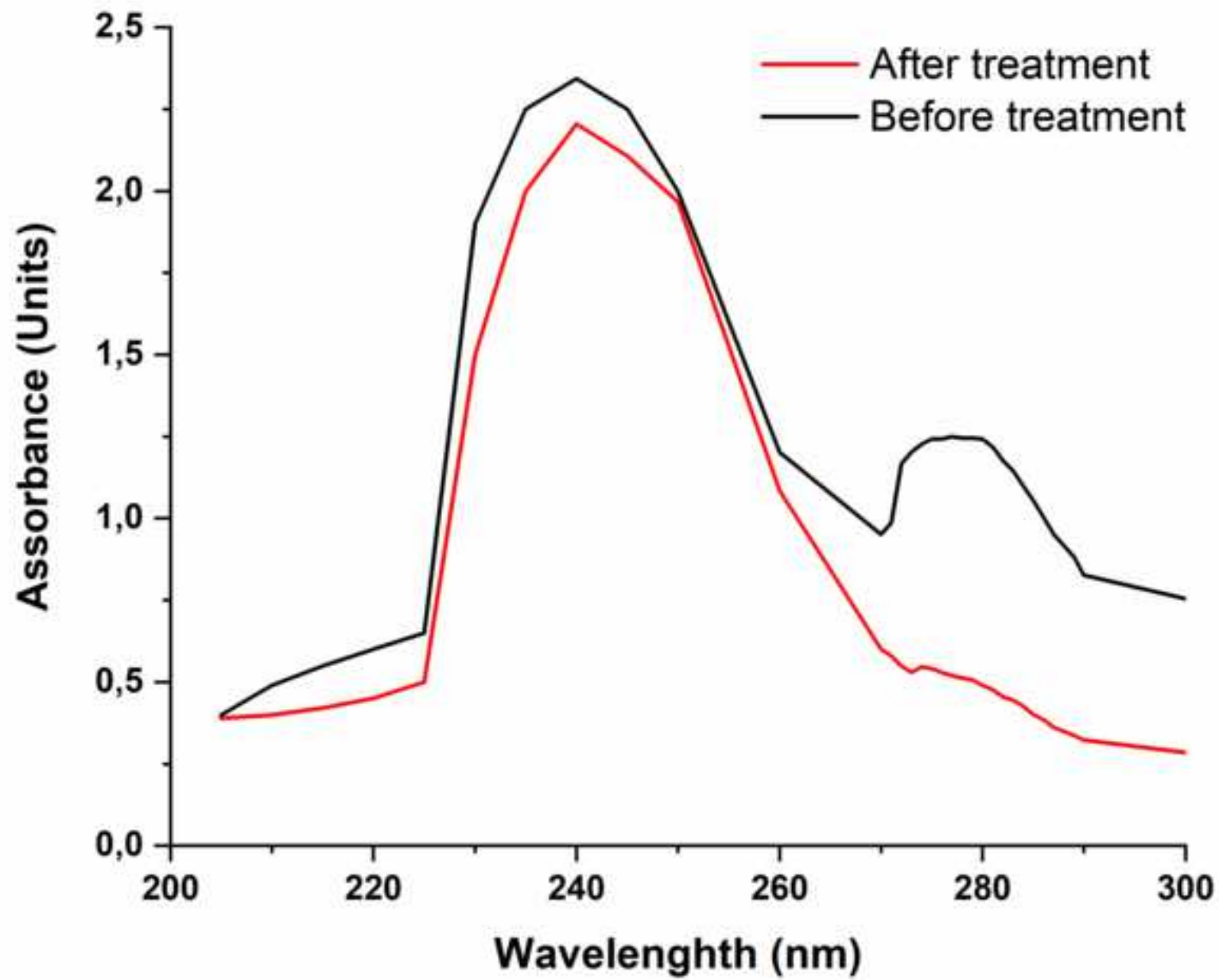

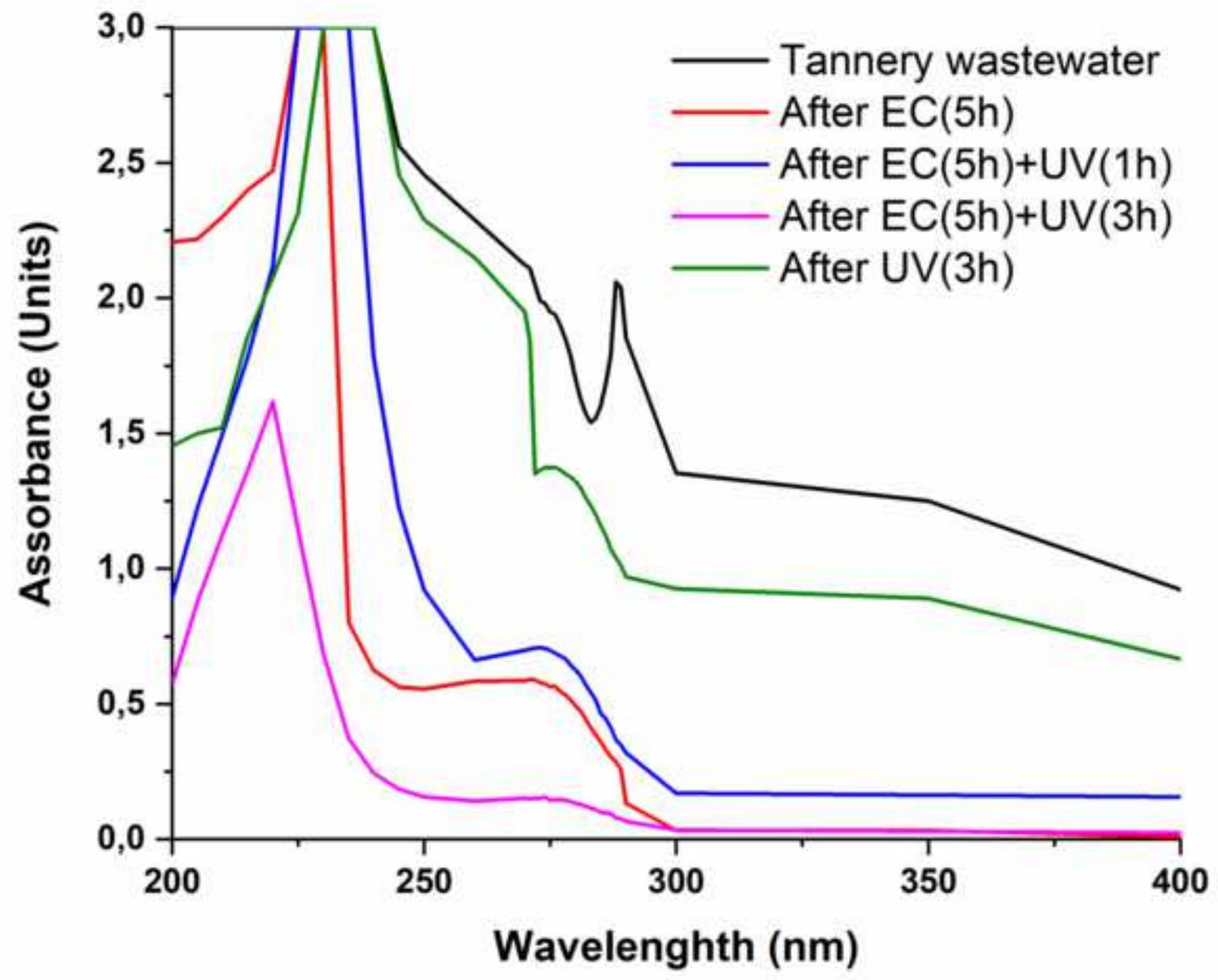

Figure 7

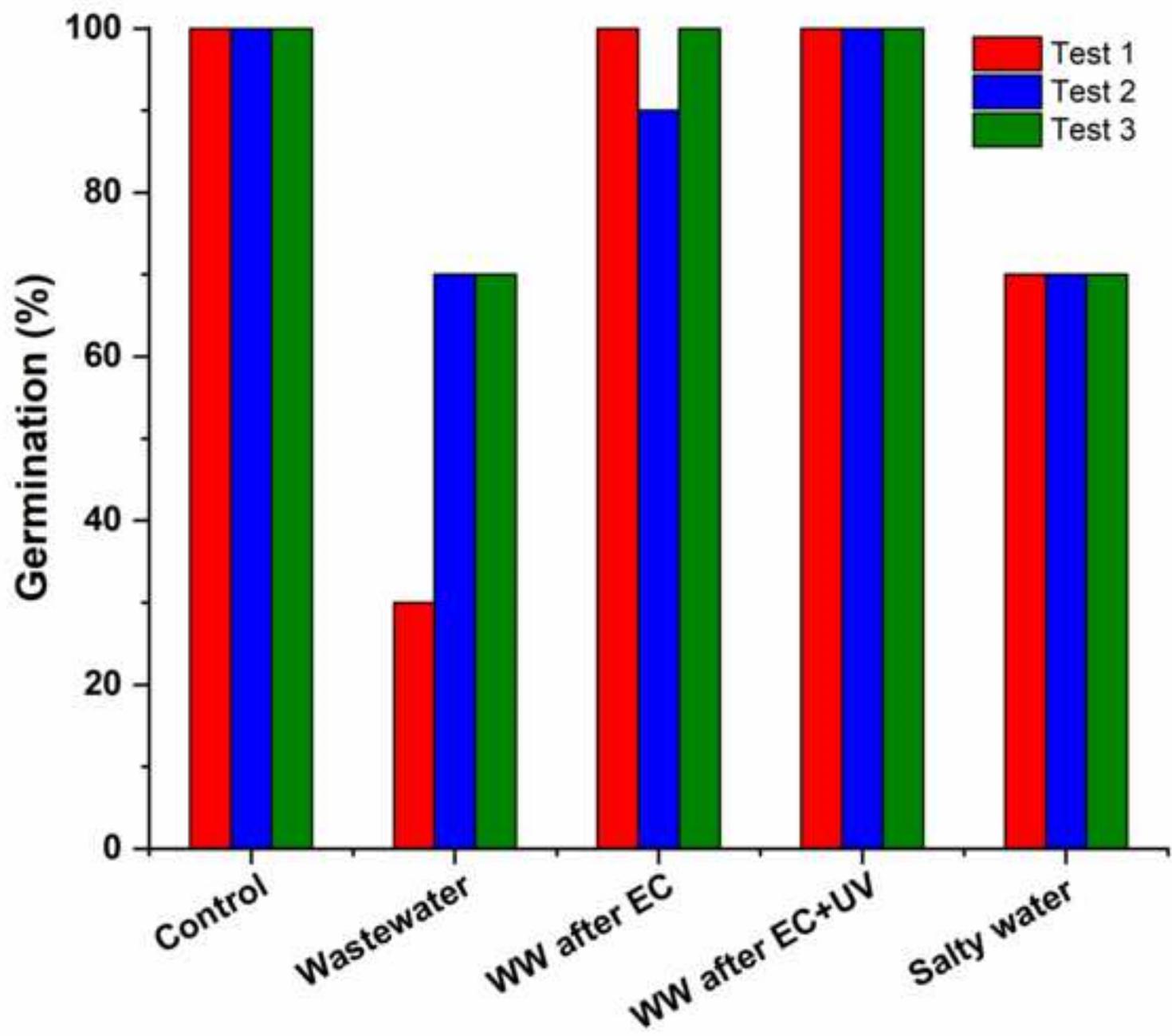

Scheme 1

a) Anode:

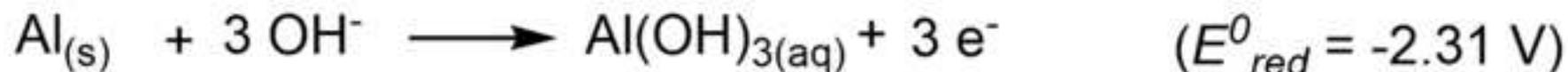
$Al_{(s)} + 3\,OH^- \longrightarrow Al(OH)_{3(aq)} + 3\,e^-$   ($E^0_{red} = -2.31\ V$)

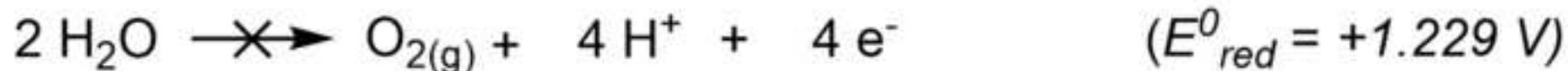
$2\,H_2O \xrightarrow{\times} O_{2(g)} + 4\,H^+ + 4\,e^-$   ($E^0_{red} = +1.229\ V$)

b) Cathode:

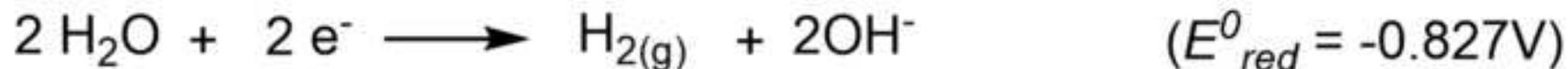
$2\,H_2O + 2\,e^- \longrightarrow H_{2(g)} + 2\,OH^-$   ($E^0_{red} = -0.827\ V$)

c) Overall redox reaction:

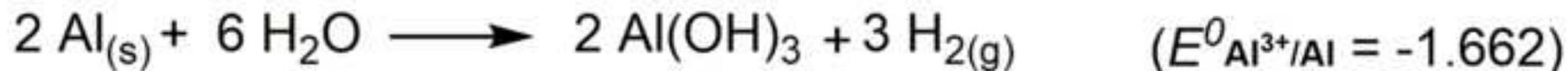
$2\,Al_{(s)} + 6\,H_2O \longrightarrow 2\,Al(OH)_3 + 3\,H_{2(g)}$   ($E^0_{Al^{3+}/Al} = -1.662$)

Scheme 2

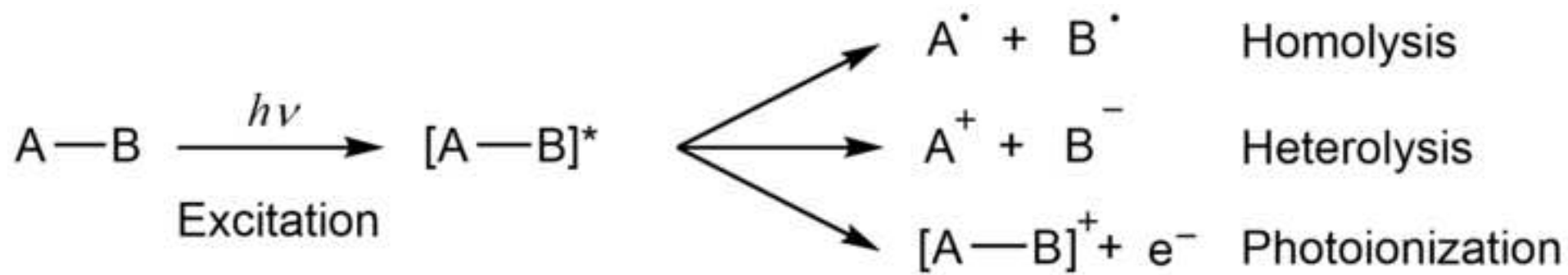